\documentclass[conference]{IEEEtran}
\usepackage{amsfonts}
\usepackage{cite,graphicx,amsmath,amsthm}
\allowdisplaybreaks[4]
\usepackage{subfigure}
\usepackage{fancyhdr}
\usepackage{dsfont}
\usepackage{array,color}
\usepackage{bm}
\usepackage{booktabs}
\usepackage{multirow}
\usepackage{algorithm}
\usepackage{algpseudocode}

\usepackage{bbm}
\usepackage{graphicx}
\def\BibTeX{{\rm B\kern-.05em{\sc i\kern-.025em b}\kern-.08em
    T\kern-.1667em\lower.7ex\hbox{E}\kern-.125emX}}

\newtheorem{theorem}{Theorem}

\newtheorem{lemma}{Lemma}

\newtheorem{proposition}{Proposition}

\newtheorem{corollary}{Corollary}

\newtheorem{property}{Property}

\newtheorem{remark}{Remark}

\newtheorem{claim}{Claim}

\begin{document}
\title{{Joint Communication and Motion Energy Minimization in UGV Backscatter Communication}}

\author{Shuai Wang$^{*}$, Minghua Xia$^{\dag}$,
        and Yik-Chung Wu$^{*}$\\
        $^{*}$Department of Electrical and Electronic Engineering, The University of Hong Kong, Hong Kong\\
        $^{\dag}$School of Electronics and Information Technology, Sun Yat-sen University, Guangzhou, 510006, China\\
E-mail: swang@eee.hku.hk; xiamingh@mail.sysu.edu.cn; ycwu@eee.hku.hk
        }

\maketitle

\vspace{-0.5in}
\begin{abstract}
While backscatter communication emerges as a promising solution to reduce power consumption at IoT devices, the transmission range of backscatter communication is short.
To this end, this work integrates unmanned ground vehicles (UGVs) into the backscatter system.
With such a scheme, the UGV could facilitate the communication by approaching various IoT devices.
However, moving also costs energy consumption and a fundamental question is: what is the right balance between spending energy on moving versus on communication?
To answer this question, this paper proposes a joint graph mobility and backscatter communication model.
With the proposed model, the total energy minimization at UGV is formulated as a mixed integer nonlinear programming (MINLP) problem.
Furthermore, an efficient algorithm that achieves a local optimal solution is derived, and it leads to automatic trade-off between spending energy on moving versus on communication.
Numerical results are provided to validate the performance of the proposed algorithm.
\end{abstract}


\section{Introduction}

With a wide range of commercial and industrial applications, Internet of Things (IoT) market is continuously growing, with the number of inter-connected IoT devices expected to exceed 20 billion by 2020 \cite{iot1}.
However, these massive IoT devices (e.g., sensors and tags) are usually limited in size and energy supply \cite{xia}, making data collection challenging in IoT systems.
To this end, backscatter communication is a promising solution, because it eliminates radio frequency (RF) components in IoT devices \cite{back1,back2}.
Unfortunately, due to the round-trip path-loss, the transmission range of backscatter communication is limited \cite{back4,back5,back6}.
This can be seen from a recent prototype in \cite{back1}, where the wirelessly powered backscatter communication only supports a range of $1$ meter at a data-rate of $1$ $\mathrm{kbps}$.

To combat the short communication range, this paper investigates a viable solution that the backscatter transmitter and receiver are mounted on an unmanned ground vehicle (UGV).
With such a scheme, the UGV could vary its location for wireless data collection, thus having the flexibility of being close to different IoT devices at different times \cite{ugv1}.
However, since moving the UGV would consume motion energy, an improperly chosen path might lead to excessive movement, thus offseting the benefit brought by movement \cite{ugv1,ugv2,ugv3}.
Therefore, the key is to balance the trade-off between spending energy on moving versus on communication, which unfortunately cannot be handled by traditional vehicle routing algorithms \cite{vr1,vr2,vr3}, since they do not take the communication power and quality-of-service (QoS) into account.

In view of the apparent research gap, this paper proposes an algorithm that leads to automatic trade-off in spending energy on moving versus on communication.
In particular, the proposed algorithm is obtained by integrating the graph mobility model and the backscatter communication model.
With the proposed model, the joint mobility management and power allocation problem is formulated as a QoS constrained energy minimization problem.
Nonetheless, such a problem turns out to be a mixed integer nonlinear programming problem (MINLP), which is nontrivial to solve due to the nonlinear coupling between discrete variables brought by moving and continuous variables brought by communication.
This is in contrast to unmanned aerial vehicle communication in which only continuous variables are involved \cite{uav}.
To this end, an efficient algorithm, which is guaranteed to obtain a local optimal solution, is proposed.
By adopting the proposed algorithm, simulation results are presented to further demonstrate the performance of the proposed algorithm under various noise power levels at IoT devices.

\emph{Notation}.
Italic letters, simple bold letters, and capital bold letters represent scalars, vectors, and matrices, respectively.
Curlicue letters represent sets and $|\cdot|$ is the cardinality of a set.
We use $(a_1,a_2,\cdots)$ to represent a sequence and $[a_1,a_2,\cdots]^{T}$ to represent a column vector, with $(\cdot)^{T}$ being the transpose operator.
The operators $\textrm{Tr}(\cdot)$ and $(\cdot)^{-1}$ take the trace and the inverse of a matrix, respectively.
Finally, $\mathbb{E}(\cdot)$ represents the expectation of a random variable.

\section{System Model}

\subsection{Mobility Model}

\begin{figure}[!t]
\centering
\includegraphics[width=50mm]{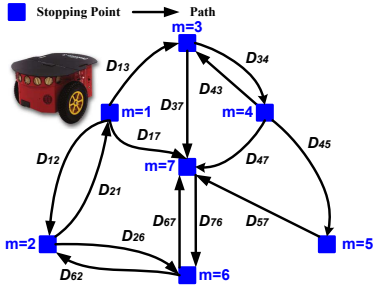}
\caption{An illustration of UGV mobility model with $M=7$.}
\label{fig_sim}
\end{figure}

\setcounter{secnumdepth}{4}We consider a wireless data collection system, which consists of $K$ IoT users and one UGV equipped with a RF transmitter and a tag reader.
The environment in which the UGV operates in is described by a directed graph $(\mathcal{V},\mathcal{E})$ as shown in Fig. 1, where $\mathcal{V}$ is the set of $M$ vertices representing the possible stopping points, and $\mathcal{E}$ is the set of directed edges representing the allowed movement paths \cite{graph}.
To quantify the path length, a matrix $\mathbf{D}=[D_{1,1},\cdots,D_{1,M};\cdots;D_{M,1},\cdots,D_{M,M}]\in\mathbb{R}^{M\times M}_+$ is defined, with the element $D_{m,j}$ representing the distance from vertex $m$ to vertex $j$ ($D_{m,m}=0$ for any $m$).
If there is no allowed path from vertex $m$ to vertex $j$, we set $D_{m,j}=+\infty$ \cite{graph}.
To model the movement of the UGV, we define a visiting path $\mathcal{Q}=\{y_1,y_2,\cdots,y_Q\}$ where $y_j \in \mathcal{V}$ for $j=1,\cdots,Q$ and $(y_j,y_{j+1})\in \mathcal{E}$ for $j=1,\cdots,Q-1$, with $Q-1$ being the number of steps to be taken.
Without loss of generality, we assume the following two conditions hold:
\begin{itemize}
\item[(i)] $y_1=y_Q$. This is generally true as a typical UGV management scenario is to have the UGV standing by at the starting point (e.g., for charging and maintenance services) after the data collection task  \cite{vr1}.
    For notational simplicity, it is assumed that vertex $y_1=y_Q=1$ is the start and end point of the path to be designed.

\item[(ii)] There are no repeating vertices among $(y_1,\cdots,y_{Q-1})$. This is true because if a vertex $m$ is visited twice, we can always introduce an auxiliary vertex with $D_{M+1,j}=D_{m,j}$ and $D_{j,M+1}=D_{j,m}$ for all $j\in\mathcal{V}$  \cite{laporte}.
    Thus this scenario can be represented by an extended graph with one more vertex and an extended $\mathbf{D}$ with dimension $(M+1)\times(M+1)$.
\end{itemize}

Correspondingly, we define the selection variable $\mathbf{v}=[v_1,\cdots,v_M]^{T}\in\{0,1\}^M$, where $v_m=1$ if the vertex $m$ appears in the path $\mathcal{Q}$ and $v_m=0$ otherwise.
Furthermore, we define a matrix $\mathbf{W}=[W_{1,1},\cdots,W_{1,M};\cdots;W_{M,1},\cdots,W_{M,M}]\in\{0,1\}^{M\times M}$, with $W_{y_j,y_{j+1}}=1$ for all $j=1,\cdots,Q-1$ and zero otherwise.

With the moving time from the vertex $m$ to the vertex $j$ being $D_{m,j}/a$ where $a$ is the velocity,
the total moving time along path $\mathcal{Q}$ is
\begin{align}
\frac{1}{a}\sum_{m=1}^M\sum_{j=1}^MW_{m,j}D_{m,j}=\frac{\mathrm{Tr}(\mathbf{D}^{T}\mathbf{W})}{a}. \label{SW}
\end{align}
Furthermore, since the total motion energy $E_M$ of the UGV is proportional to the total motion time \cite{ugv1,ugv2,ugv3}, the motion energy can be expressed in the form of
\begin{align}
E_M=\left(\frac{\alpha_1}{a}+\alpha_2\right)\mathrm{Tr}(\mathbf{D}^{T}\mathbf{W}), \label{EW}
\end{align}
where $\alpha_1$ and $\alpha_2$ are parameters of the model (e.g., for a Pioneer 3DX robot in Fig. 1, $\alpha_1=0.29$ and $\alpha_2=7.4$ \cite[Sec. IV-C]{ugv1}).

\subsection{Backscatter Communication Model}

Based on the mobility model, the UGV moves along the selected path $\mathcal{Q}$ to collect data from users.
In particular, from the starting point $y_1$, the UGV stops for a duration $u_{y_1}$ and then it moves along edge $(y_1,y_2)$ to its outward neighbor $y_2$, and stops for a duration $u_{y_2}$. The UGV keeps on moving and stopping along the path until it reaches the destination $y_Q$.

When the UGV stops at the vertex $m$ (with $v_m=1$), it will wait for a time duration $u_m$ for data collection.
Out of this $u_m$, a duration of $t_{k,m}$ will be assigned to collect data from user $k$ via full-duplex backscatter communication\footnote{When user $k$ adapts the variable impedance for modulating the backscattered waveform, other users keep silent to avoid collision \cite{back4}.} \cite{back2}. More specifically, if $t_{k,m}=0$, the IoT user $k$ will not be served in duration $u_m$.
On the other hand, if $t_{k,m}\neq0$,the RF source at the UGV transmits a symbol $x_{k,m}\in\mathbb{C}$ with $\mathbb{E}[|x_{k,m}|^2]=p_{k,m}$, where $p_{k,m}$ is the transmit power of the RF source.
Then the received signal-to-noise ratio (SNR) at the UGV tag reader is
$\eta |g_{k,m}|^2|h_{k,m}|^2 p_{k,m}/N_0$, where $h_{k,m}\in \mathbb{C}$ is the downlink channel from the UGV to user $k$, $g_{k,m}\in \mathbb{C}$ is the uplink channel\footnote{If the environment is static, ray tracing methods \cite{E1} could be used to estimate $\{g_{k,m},h_{k,m}\}$.
On the other hand, if the channel is varying but with a fixed distribution, we could allow the UGV to collect a small number of measurements at the stopping points before a set of new missions (e.g., three to five missions) \cite{E2}, and then the UGV can predict $\{g_{k,m},h_{k,m}\}$.} from user $k$ to the UGV, and $N_0$ is the power of complex Gaussian noise (including the self-interference due to full-duplex communication \cite{liwang,zwen,zwen2}).
Furthermore, $\eta$ is the tag scattering efficiency determined by the load impedance and the antenna impedance \cite{back10}.

Based on the backscatter model, the transmission rate during $t_{k,m}$ is given by
\begin{align}\label{rate}
R_{k,m}=\mathrm{log}_2\left(1+v_m\cdot\frac{\beta\eta|g_{k,m}|^2|h_{k,m}|^2 p_{k,m}}{N_0}\right),
\end{align}
where $\beta$ is the performance loss due to imperfect modulation and coding schemes in backscatter communication \cite{back11}.
For example, in bistatic backscatter communication with frequency shift keying, $\beta=0.5$ \cite{back11}.
On the other hand, in ambient backscatter communication with on-off keying, $\beta$ is obtained by fitting the logarithm function $\mathrm{log}_2\left(1+\beta x\right)$ to $1-\mathbb{Q}\left(\sqrt{x}\right)$ \cite{ook}, where
$\mathbb{Q}\left(x\right)=1/\sqrt{2\pi}\int_x^\infty \mathrm{exp}\left(-u^2/2 \right)\mathrm{d}u$ refers to the Q-function.

\section{Joint Communication and Motion Energy Miminization}

In wireless data collection systems, the task is to collect certain amount of data from different IoT devices by planning the path (involving variables $\mathbf{v}$ and $\mathbf{W}$) and designing the stopping time $\{t_{k,m}\}$ and transmit power $\{p_{k,m}\}$.
In particular, the data collection QoS requirement of the $k^{\mathrm{th}}$ IoT device can be described by
\begin{align}
&\sum_{m=1}^M
t_{k,m}\cdot\mathrm{log}_2\left(1+v_m\cdot\frac{\beta\eta|g_{k,m}|^2|h_{k,m}|^2 p_{k,m}}{N_0}\right)\geq\gamma_k,
\end{align}
where $\gamma_k>0$ (in $\mathrm{bit/Hz}$) is the amount of data to be collected from user $k$.

Notice that the variables $\mathbf{v}$ and $\mathbf{W}$ are dependent since $v_m=0$ implies $W_{m,j}=W_{j,m}=0$ for any $j\in \mathcal{V}$.
On the other hand, the UGV would visit the vertex with $v_m=1$, making $\sum_{j=1}^M W_{m,j} = \sum_{j=1}^M W_{j,m}=1$.
Combining the above two cases, we have
\begin{align}
&\sum_{j=1}^MW_{m,j}=v_m,~\sum_{j=1}^MW_{j,m}=v_m,~~\forall m=1,\cdots,M.
\end{align}
Furthermore, since the path must be connected, the following subtour elimination constraints are required to eliminate disjointed sub-tours \cite{laporte}:
\begin{align}
&\lambda_m-\lambda_j+\left(\sum_{l=1}^Mv_l-1\right)W_{m,j}+\left(\sum_{l=1}^Mv_l-3\right)W_{j,m}
\nonumber\\
&\leq \sum_{l=1}^Mv_l-2
+J\left(2-v_m-v_j\right)
,~~\forall m,j\geq 2,~m\neq j,
\nonumber\\
&v_m\leq\lambda_m\leq\left(\sum_{l=1}^Mv_l-1\right)v_m,~~\forall m\geq 2,
\end{align}
where $\{\lambda_{m}\}$ are slack variables to guarantee a connected path, and $\sum_{l=1}^Mv_l$ is the number of vertices involved in the path.
The constant $J=10^{6}$ is large enough such that the first line of constraint is always satisfied when $v_m=0$ or $v_j=0$.
In this way, the vertices not to be visited would not participate in subtour elimination constraints.

Having the data collection and graph mobility constraints satisfied, it is then crucial to reduce the total energy consumption at the UGV.
As the energy consumption includes motion energy
$E_M=\left(\alpha_1/a+\alpha_2\right)\mathrm{Tr}(\mathbf{D}^{T}\mathbf{W})$
and communication energy $E_C=\sum_{m=1}^M\sum_{k=1}^Kt_{k,m}p_{k,m}$, the joint mobility management and power allocation problem of the data collection system is formulated as $\mathrm{P}1$, where \eqref{P1b} is for constraining the operation (including moving and data collection) to be completed within $T$ seconds, and \eqref{notvisit} is for constraining the stopping time to be zero if the vertex is not visited.
It can be seen from the constraint \eqref{P1a} of $\mathrm{P}1$ that the UGV can choose the stopping vertices, which in turn affect the channel gains to and from the IoT users.
By choosing the stopping vertices with better channel gains to IoT users, the transmit powers $\{p_{k,m}\}$ might be reduced.
However, this might also lead to additional motion energy, which in turn costs more energy consumption at the UGV.
Therefore, there exists a trade-off between moving and communication, and solving $\mathrm{P}1$ can concisely balance this energy trade-off.

Unfortunately, problem $\mathrm{P}1$ is nontrivial to solve due to the following reasons.
Firstly, $\mathrm{P}1$ is NP-hard, since it involves the integer constraints \eqref{edge}$-$\eqref{vertex} \cite{dimitri}.
Secondly, the data-rate and the energy cost at each vertex are dependent on the transmit power $\{p_{k,m}\}$ and transmission time $\{t_{k,m}\}$, which are unknown.
This is in contrast to traditional integer programming problems \cite{dimitri}, where the reward of visiting each vertex is a constant.
\begin{subequations}
\begin{align}
\mathrm{P}1:
&\mathop{\mathrm{min}}_{\substack{\mathbf{v},\mathbf{W},\{\lambda_m\}\\ \{t_{k,m},p_{k,m}\}}}
~\left(\frac{\alpha_1}{a}+\alpha_2\right)\mathrm{Tr}(\mathbf{D}^{T}\mathbf{W})+\sum_{m=1}^M\sum_{k=1}^Kt_{k,m}p_{k,m}
\nonumber
\\
\mathrm{s.t.}~&\sum_{m=1}^M
t_{k,m}\cdot\mathrm{log}_2\left(1+v_m\cdot\frac{\beta\eta|g_{k,m}|^2|h_{k,m}|^2 p_{k,m}}{N_0}\right)
\nonumber\\
&\geq\gamma_k,~~\forall k, \label{P1a}
\\
&\frac{1}{a}\mathrm{Tr}(\mathbf{D}^{T}\mathbf{W})+\sum_{m=1}^M\sum_{k=1}^Kt_{k,m}\leq T, \label{P1b}
\\
&
\sum_{j=1}^MW_{m,j}=v_m,~\sum_{j=1}^MW_{j,m}=v_m,~~\forall m, \label{P1c}
\\
&\lambda_m-\lambda_j+\left(\sum_{l=1}^Mv_l-1\right)W_{m,j}+\left(\sum_{l=1}^Mv_l-3\right)W_{j,m}
\nonumber\\
&
\leq \sum_{l=1}^Mv_l-2+J\left(2-v_m-v_j\right)
,
\nonumber\\
&
\forall m,j\geq 2,~m\neq j, \label{subtour1}
\\
&v_m\leq\lambda_m\leq\left(\sum_{l=1}^Mv_l-1\right)v_m,~~\forall m\geq 2, \label{subtour2}
\\
&W_{m,j}\in\{0,1\},~~\forall m,j,~~W_{m,m}=0,~~\forall m, \label{edge}
\\
&v_{1}=1,~v_{m}\in\{0,1\},~~\forall m\geq 2, \label{vertex}
\\
&(1-v_m)\cdot t_{k,m}=0,~~\forall k,m, \label{notvisit}
\\
&t_{k,m}\geq0,~p_{k,m}\geq0,~~\forall k,m, \label{resource}
\end{align}
\end{subequations}

\section{Local Optimal Solution to $\mathrm{P}1$}

Despite the optimization challenges, this section proposes an algorithm that theoretically achieves a local optimal solution to $\mathrm{P}1$.
The insight behind this algorithm is to derive the optimal solution of $\mathbf{W},\{\lambda_m\},\{t_{k,m},p_{k,m}\}$ to $\mathrm{P}1$ with fixed $\mathbf{v}$.
By representing $\mathbf{W},\{\lambda_m\},\{t_{k,m},p_{k,m}\}$  as functions of $\mathbf{v}$, problem $\mathrm{P}1$ is simplified to an equivalent problem only involving $\mathbf{v}$.
Then we will can capitalize on the successive local search (SLS) method \cite{meta,sls1,sls2} to obtain the local optimal solution.

\subsection{Optimal Solution of $\mathbf{W}$ and $\{t_{k,m},p_{k,m}\}$  with Fixed $\mathbf{v}$}

When $\mathbf{v}=\widetilde{\mathbf{v}}$, where $\widetilde{\mathbf{v}}$ is any feasible solution to $\mathrm{P}1$, the constraint \eqref{vertex} can be dropped since it only involves $\mathbf{v}$.
Moreover, to resolve the nonlinear coupling between $\{t_{k,m}\}$ and $\{p_{k,m}\}$, we replace $\{p_{k,m}\}$ with a new variable $\{Q_{k,m}\}$ such that
$\{Q_{k,m}:=t_{k,m}p_{k,m}\}$.
Based on the above variable substitution, the objective function of $\mathrm{P}1$ becomes
\begin{align}
&\left(\alpha_1/a+\alpha_2\right)\mathrm{Tr}(\mathbf{D}^{T}\mathbf{W})+\sum_{m=1}^M\sum_{k=1}^KQ_{k,m}, \label{obj}
\end{align}
which is linear.
On the other hand, the constraint \eqref{P1a} is equivalent to
\begin{align}\label{P1a2}
&\sum_{m=1}^M\underbrace{
t_{k,m}\cdot\mathrm{log}_2\left(1+\frac{A_{k,m} Q_{k,m}}{t_{k,m}}\right)}
_{:=\Phi_{k,m}(t_{k,m},Q_{k,m})}
\geq\gamma_k,~~\forall k,
\end{align}
where the constant
\begin{align}\label{Akm}
A_{k,m}:=\widetilde{v}_m\cdot\frac{\beta\eta|g_{k,m}|^2|h_{k,m}|^2}{N_0},
\end{align}
and the following property can be established.
\begin{property}
(i) The function $\Phi_{k,m}$ is concave with respect to $\{t_{k,m},Q_{k,m}\}$.
(ii) $\Phi_{k,m}$ is a monotonically increasing function of $t_{k,m}$ for all $(k,m)$.
\end{property}
\begin{proof}
To prove part (i) of this property, we note that $\Phi_{k,m}$ is the perspective transformation of the concave function $\mathrm{log}_2\left(1+A_{k,m} Q_{k,m}\right)$.
Since perspective transformation preserves concavity \cite{opt1}, $\Phi_{k,m}$ is also concave.

To prove part (ii), we compute the derivative of $\Phi_{k,m}$ in \eqref{P1a2} with respect to $t_{k,m}$ as
\begin{align}
\nabla_{t_{k,m}}\Phi_{k,m}=&
\mathrm{log}_2\left(1+\frac{A_{k,m} Q_{k,m}}{t_{k,m}}\right)
\nonumber\\
&-\frac{1}{\mathrm{ln}2}\cdot \frac{A_{k,m}Q_{k,m}}{t_{k,m}+A_{k,m}Q_{k,m}}.
\end{align}
Using the result from part (i), we have $\nabla^2_{t_{k,m}}\Phi_{k,m}\leq 0$ due to $\Phi_{k,m}$ being concave.
Therefore, $\nabla_{t_{k,m}}\Phi_{k,m}$ is a monotonically decreasing function of $t_{k,m}$.
This means that
\begin{align}
\nabla_{t_{k,m}}\Phi_{k,m}\geq \mathop{\mathrm{lim}}_{t_{k,m}\rightarrow+\infty} \nabla_{t_{k,m}}\Phi_{k,m}=0,
\end{align}
and the proof is completed.
\end{proof}

Based on the result from part (i) of \textbf{Property 1}, it is clear that the constraint \eqref{P1a2} is convex.
On the other hand, according to part (ii) of \textbf{Property 1}, it can be seen that the optimal $\mathbf{W}^*$ and $\{t_{k,m}^*\}$ to $\mathrm{P}1$ must activate the constraint \eqref{P1b}.
Otherwise, we can always increase the value of $\{t_{k,m}\}$ such that the left hand side of the constraint \eqref{P1a2} is increased.
This allows us to decrease the value of $\{Q_{k,m}\}$ (thus the objective value of \eqref{obj}), which contradicts to $\{t_{k,m}^*\}$ being optimal.
As a result, the constraint \eqref{P1b} can be restricted into an equality $\mathrm{Tr}(\mathbf{D}^{T}\mathbf{W})/a+\sum_{m=1}^M\sum_{k=1}^Kt_{k,m}=T$, giving
\begin{align}\label{DW}
\mathrm{Tr}(\mathbf{D}^{T}\mathbf{W})=a\left(T-\sum_{m=1}^M\sum_{k=1}^Kt_{k,m}\right).
\end{align}
Putting \eqref{DW} into \eqref{obj}, $\mathrm{P}1$ is equivalently transformed into the following two-stage optimization problem:
\begin{align}
\mathrm{P}2:
\mathop{\mathrm{min}}_{\substack{\{t_{k,m},Q_{k,m}\}}}
~&\left(\alpha_1+\alpha_2a\right)\left(T-\sum_{m=1}^M\sum_{k=1}^Kt_{k,m}\right)
\nonumber\\
&
+\sum_{m=1}^M\sum_{k=1}^KQ_{k,m}
\nonumber\\
\mathrm{s.t.}~~~~~&\sum_{m=1}^M\Phi_{k,m}(t_{k,m},Q_{k,m})
\geq\gamma_k,~~\forall k,
\nonumber\\
&
\sum_{m=1}^M\sum_{k=1}^Kt_{k,m}
=\mathop{\mathrm{max}}_{\substack{\{\mathbf{W},\lambda_m\}}} \Big\{T-\frac{\mathrm{Tr}(\mathbf{D}^{T}\mathbf{W})}{a}:
\nonumber\\
&~~~~~~~~~~~~~~~~~~~~
\eqref{P1c}-\eqref{edge}\Big\},\nonumber
\\
&(1-\widetilde{v}_m)\cdot t_{k,m}=0,~~\forall k,m, \nonumber
\\
&t_{k,m}\geq0,~Q_{k,m}\geq0,~~\forall k,m.
\end{align}
To solve $\mathrm{P}2$, we first need to compute the right hand side of the second constraint, which leads to the following problem:
\begin{align}
&\mathop{\mathrm{max}}_{\substack{\mathbf{W},\{\lambda_{m}\}}}
~T-\frac{\mathrm{Tr}(\mathbf{D}^{T}\mathbf{W})}{a}
~~~\mathrm{s.t.}~\eqref{P1c}-\eqref{edge}. \label{TSP}
\end{align}
The problem \eqref{TSP} is a travelling salesman problem, which can be optimally solved via the software Mosek \cite{laporte,MINLP}.
Denoting the optimal solution to the problem \eqref{TSP} as $\{\widehat{\mathbf{W}},\widehat{\lambda}_m\}$,
the optimal objective value of the travelling salesman problem is given by
$\Upsilon(\widetilde{\mathbf{v}}):=
T-\mathrm{Tr}(\mathbf{D}^{T}\widehat{\mathbf{W}})/a$.
Finally, by putting the obtained $\Upsilon(\widetilde{\mathbf{v}})$ into $\mathrm{P}2$, the second constraint of $\mathrm{P}2$ is written as $\sum_{m=1}^M\sum_{k=1}^Kt_{k,m}
= \Upsilon(\widetilde{\mathbf{v}})$.
Adding to the fact that all the other constraints in $\mathrm{P}2$ are convex, $\mathrm{P}2$ is a convex optimization problem.
Therefore, $\mathrm{P}2$ can be optimally solved by CVX, a Matlab software for solving convex problems \cite{opt1}.
Denoting its solution as $\{\widehat{t}_{k,m},\widehat{Q}_{k,m}\}$, the optimal $\{\widehat{p}_{k,m}\}$ with fixed $\mathbf{v}=\widetilde{\mathbf{v}}$ can be recovered as $\widehat{p}_{k,m}=\widehat{Q}_{k,m}/\widehat{t}_{k,m}$.

\subsection{Local Optimal Solution of $\mathbf{v}$}

With path selection $\widehat{\mathbf{W}}$, transmit times $\{\widehat{t}_{k,m}\}$, and transmit powers $\{\widehat{p}_{k,m}\}$ derived in Section IV-A, the optimal objective value of $\mathrm{P}1$ with $\mathbf{v}=\widetilde{\mathbf{v}}$ can be written as
\begin{align}
&\Xi(\widetilde{\mathbf{v}})=
\left(\alpha_1/a+\alpha_2\right)\mathrm{Tr}(\mathbf{D}^{T}\widehat{\mathbf{W}})
+\sum_{m=1}^M\sum_{k=1}^K\widehat{t}_{k,m}\widehat{p}_{k,m}.
\end{align}
Therefore, problem $\mathrm{P}1$ is re-written as
\begin{align}
&\mathrm{P}3:\mathop{\mathrm{min}}_{\substack{\mathbf{v}}}
~~\Xi(\mathbf{v})
~~\mathrm{s.t.}~~v_{1}=1,~v_{m}\in\{0,1\},~~\forall m\geq 2. \label{v}
\end{align}
To solve $\mathrm{P}3$, a naive way is to apply exhaustive search for $\mathbf{v}$.
Unfortunately, since the searching space of $\{v_m\}$ is very large (i.e., $2^{M-1}$), direct implementation of exhaustive search is impossible.
To address the above issue, a SLS method \cite{meta,sls1,sls2} is presented, which significantly reduces the computational complexity compared to exhaustive search.

More specifically, we start from a feasible solution of $\mathbf{v}$ (e.g., $\mathbf{v}^{[0]}=[1,0,\cdots,0]^{T}$), and randomly selects a candidate solution $\mathbf{v}'$ from the neighborhood $\mathcal{N}(\mathbf{v}^{[0]})$. Since a natural neighborhood operator for binary optimization problem is to flip the values of $\{v_m\}$, $\mathcal{N}(\mathbf{v}^{[0]})$ can be set to
\begin{align}
\mathcal{N}(\mathbf{v}^{[0]})=\{\mathbf{v}\in\{0,1\}^M:||\mathbf{v}-\mathbf{v}^{[0]}||_0\leq L,~v_1=1\},
\end{align}
where $L\geq 1$ is the size of neighborhood \cite{sls2}.
It can be seen that $\mathcal{N}(\mathbf{v}^{[0]})$ is a subset of the entire feasible space and containing solutions ``close'' to $\mathbf{v}^{[0]}$.

With the neighborhood $\mathcal{N}(\mathbf{v}^{[0]})$ defined above and the choice of $\mathbf{v}$ fixed to $\mathbf{v}=\mathbf{v}'$, we consider two cases.
\begin{itemize}
\item[(i)] If $\Xi(\mathbf{v}')\leq\Xi(\mathbf{v}^{[0]})$, we update $\mathbf{v}^{[1]}\leftarrow\mathbf{v}'$. By treating $\mathbf{v}^{[1]}$ as a new feasible solution, we can construct the next neighborhood $\mathcal{N}(\mathbf{v}^{[1]})$.

\item[(ii)] If $\Xi(\mathbf{v}')>\Xi(\mathbf{v}^{[0]})$, we find another point within the neighborhood $\mathcal{N}(\mathbf{v}^{[0]})$ until $\Xi(\mathbf{v}')\leq\Xi(\mathbf{v}^{[0]})$.

\end{itemize}
The above procedure is repeated to iteratively generate a sequence of $\{\mathbf{v}^{[1]},\mathbf{v}^{[2]},\cdots\}$ and the converged point is guaranteed to be a local optimal solution to $\mathrm{P}1$ \cite{dimitri}.
In practice, we terminate the iterative procedure when the number of iterations is larger than $\overline{\mathrm{Iter}}$.

\subsection{Summary of Algorithm}

Since the Algorithm 1 finds the local optimal solution of $\mathbf{v}$ to $\mathrm{P}3$, and the optimal solution of $\mathbf{W}$ and $\{t_{k,m},p_{k,m}\}$ with fixed $\mathbf{v}$ can be computed according to Section IV-A, the entire algorithm for computing the local optimal solution to $\mathrm{P}3$ (equivalently $\mathrm{P}1$) is summarized in Algorithm 1.
In terms of computational complexity, computing $\Upsilon(\mathbf{v}')$ would involve the travelling salesman problem, which requires a complexity of $O\left((M-1)^2\cdot2^{M-1}\right)$ in the worst case \cite{tsp}.
On the other hand, since $\mathrm{P}2$ has $2KM$ variables, solving $\mathrm{P}2$ via CVX requires a complexity of $O\left((2KM)^{3.5}\right)$ \cite{opt2}.
Therefore, with $\overline{\mathrm{Iter}}$ iterations, the proposed Algorithm 1 requires a complexity of $O\left(\overline{\mathrm{Iter}}\left[(M-1)^2\cdot2^{M-1}+(2KM)^{3.5}\right]\right)$.

\begin{algorithm}
    \caption{Proposed local optimal solution to $\mathrm{P}1$}
        \begin{algorithmic}[1]
            \State \textbf{Initialize} $\mathbf{v}^{[0]}=[1,0,0,\cdots]^{T}$ and a proper $L$. Set counter $n=0$ and the number of iterations $\mathrm{Iter}=0$.
            \State \textbf{Repeat}
            \State \ \ Sample a solution $\mathbf{v}'\in\mathcal{N}(\mathbf{v}^{[n]})$.
            \State \ \ Compute $\Xi(\mathbf{v}')$ by solving $\mathrm{P}1$ with $\mathbf{v}=\mathbf{v}'$.
            \State \ \ If $\Xi(\mathbf{v}')\leq\Xi(\mathbf{v}^{[n]})$, update $\mathbf{v}^{[n+1]}\leftarrow\mathbf{v}'$ and $n\leftarrow n+1$.
             \State \ \ Update $\mathrm{Iter}\leftarrow \mathrm{Iter}+1$.
            \State \textbf{Until} $\mathrm{Iter}= \overline{\mathrm{Iter}}$.
            \State Output $\mathbf{v}^{[n]}$, $\widehat{\mathbf{W}}$, and $\{\widehat{t}_{k,m},\widehat{p}_{k,m}\}$.
        \end{algorithmic}
\end{algorithm}

\section{Simulation Results and Discussions}

This section provides numerical results to evaluate the performance of the UGV backscatter communication network.
It is assumed that the backscattering efficiency is $\eta=0.78$ (corresponding to $1.1~\mathrm{dB}$ loss \cite{back11}), and the performance loss due to imperfect modulation is $\beta=0.5$ \cite{back11}.
Within the time budget $T=50~\mathrm{s}$, the data collection targets $\gamma_k\sim \mathcal{U}(2,4)$ in the unit of $\mathrm{bit/Hz}$ are requested by $K=10$ IoT users (corresponding to $K\gamma_k/T=0.4\sim0.8~\mathrm{bps/Hz}$ for the system \cite{iot1}), where $\mathcal{U}(a,b)$ represents the uniform distribution within the interval $[a,b]$.

Based on the above settings, we simulate the data collection map in a $20~\mathrm{m}\times20~\mathrm{m}=400~\mathrm{m}^2$ square area, which is a typical size for smart warehouses.
Inside this map, $K=10$ IoT users and $M=15$ vertices representing stopping points are uniformly scattered.
Among all the vertices, the vertex $m=1$ is selected as the starting point of the UGV.
With the locations of all the stopping points and the IoT devices, the distances between each pair of IoT device and stopping point can be computed, and the distance-dependent path-loss model $\varrho_{k,m}=\varrho_0\cdot(\frac{d_{k,m}}{d_0})^{-2.5}$ is adopted \cite{channel}, where $d_{k,m}$ is the distance from user $k$ to the stopping point $m$, and $\varrho_0=10^{-3}$ is the path-loss at distance $d_0=1~\mathrm{m}$.
Based on the path-loss model, channels $g_{k,m}$ and $h_{k,m}$ are generated according to $\mathcal{CN}(0,\varrho_{k,m})$.
Each point in the figures is obtained by averaging over $100$ simulation runs, with independent channels and realizations of locations of vertices and users in each run.

\begin{figure}
 \centering
  \subfigure[]{
    \label{fig:subfig:b} 
    \includegraphics[width=40mm]{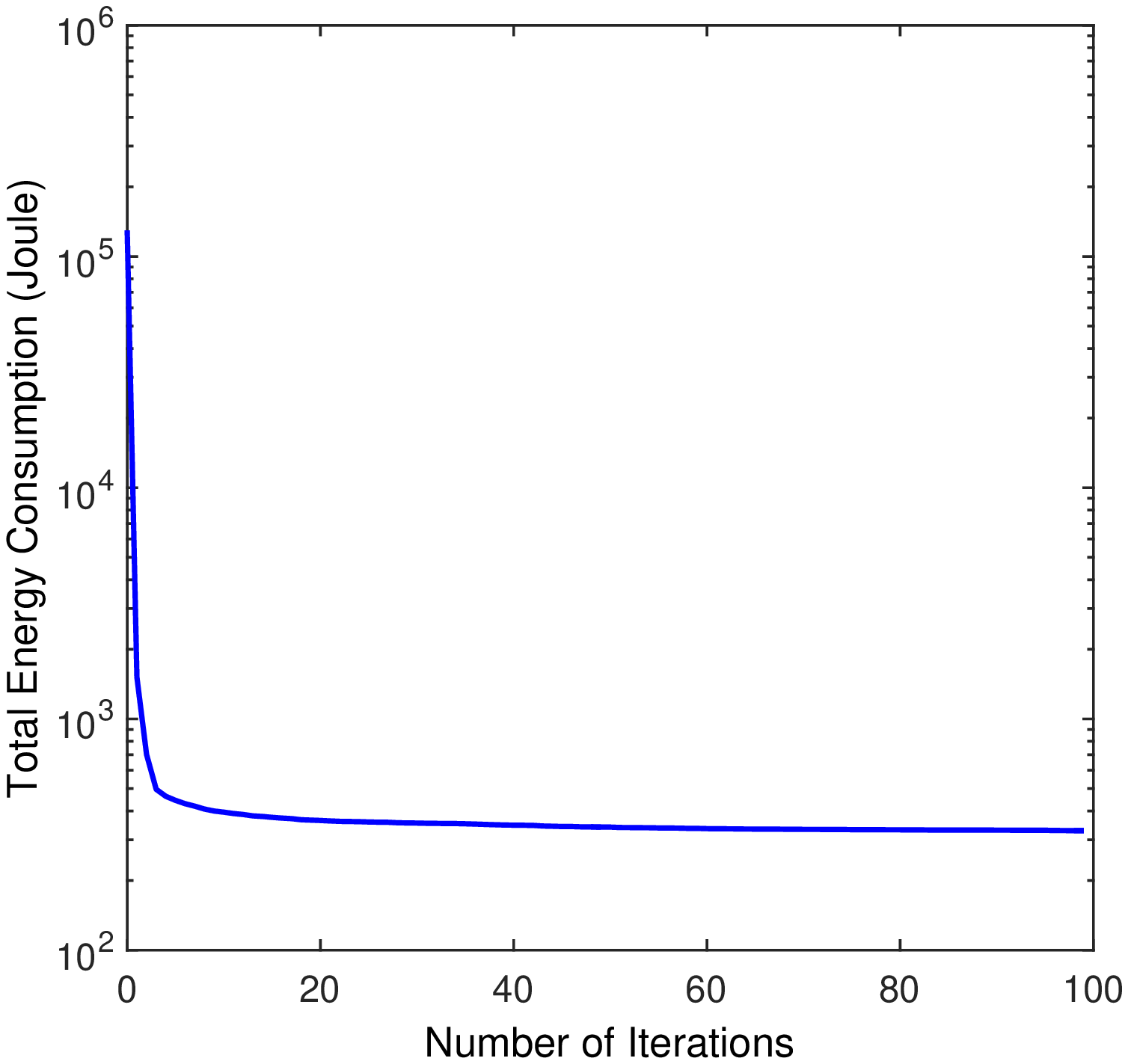}}
          \hspace{0.02in}
      \subfigure[]{
    \label{fig:subfig:b} 
    \includegraphics[width=40mm]{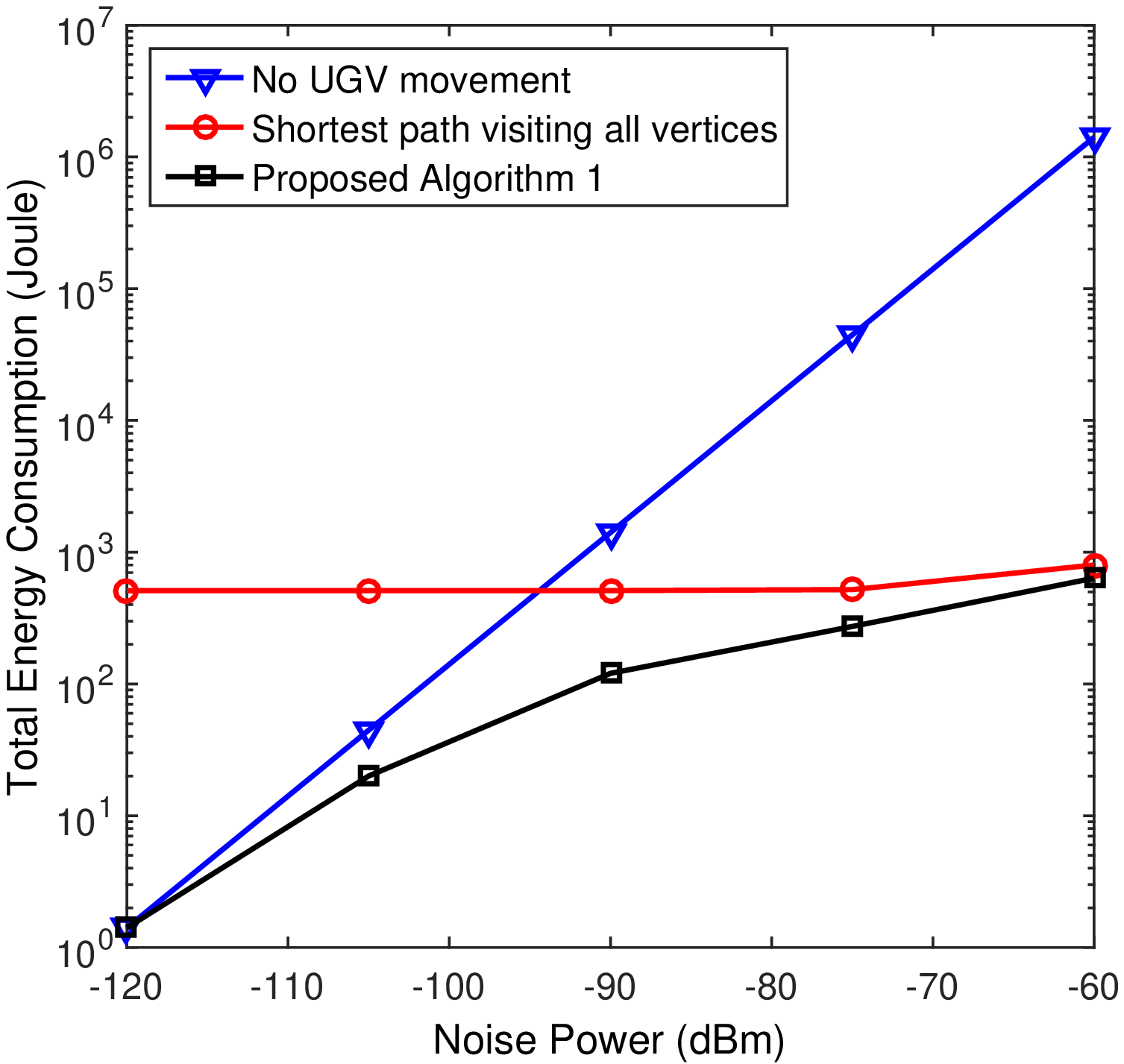}}
  \caption{(a) Total energy consumption versus the number of iterations for the case of $K=10$ and $M=15$; (b) Total energy consumption versus noise power $N_0$ with $K=10$ and $M=15$ when $\gamma_k\sim \mathcal{U}(2,4)$.
}
  \label{fig:subfig} 
\end{figure}

To verify the convergence of Algorithm 1 in Section IV, Fig. 2a shows the total energy consumption versus the number of iterations $\mathrm{Iter}$ when the receiver noise power $N_0=-70~\mathrm{dBm}$  (corresponding to power spectral density $-120~\mathrm{dBm/Hz}$ \cite{noise} with $100~\mathrm{kHz}$ bandwidth \cite{iot1}).
It can be seen that with the choice of $L=3$, the total energy consumption in the unit of joule converges and stabilizes after $50$ iterations.
This verifies the convergence property of SLS and also indicates that the number of iterations for SLS to converge is moderate.
Therefore, we set $L = 3$ with the number of iterations being 50 in the subsequent simulations.

\begin{figure}
 \centering
  \subfigure[]{
    \label{fig:subfig:b} 
    \includegraphics[width=40mm]{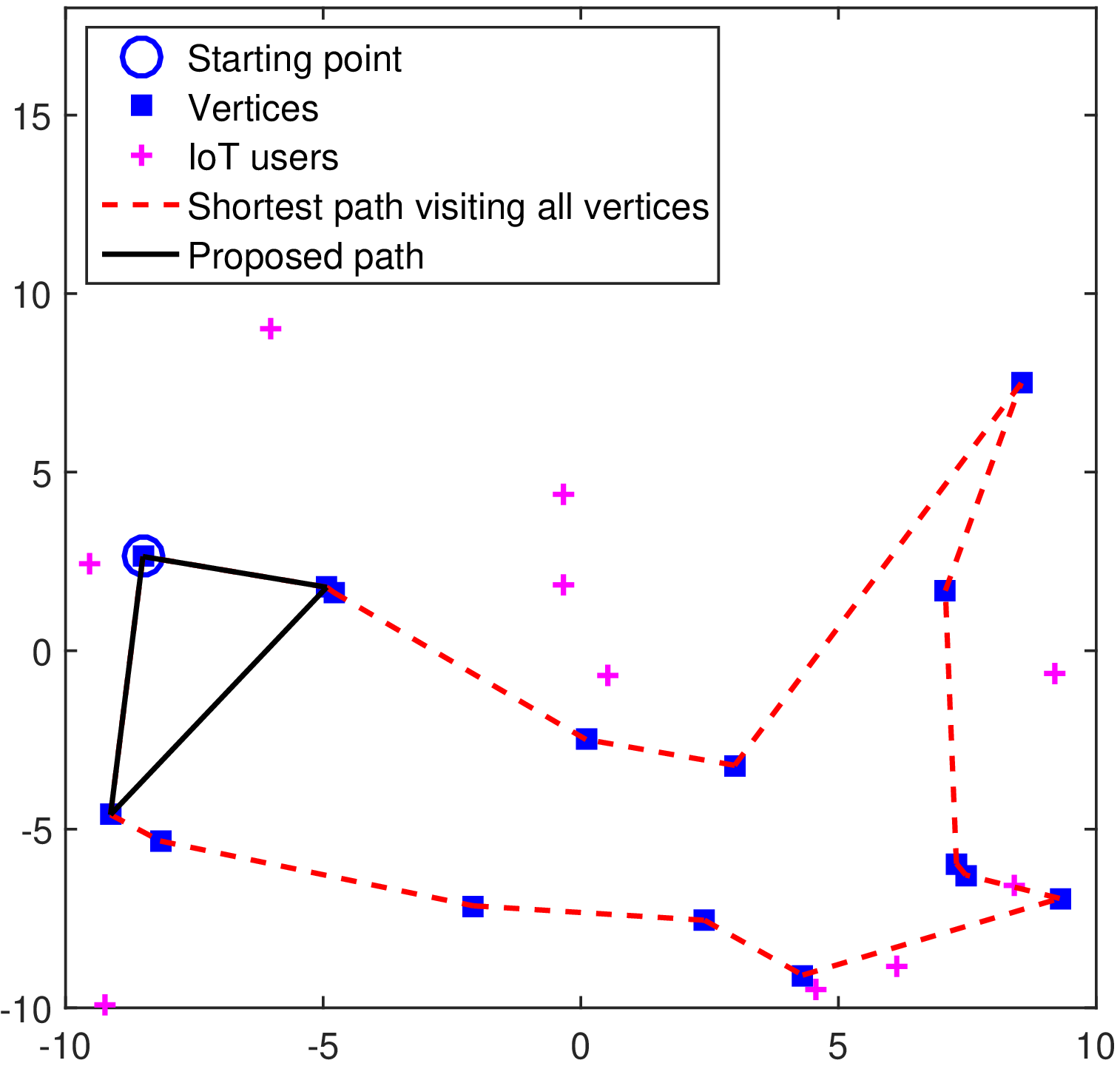}}
          \hspace{0.02in}
      \subfigure[]{
    \label{fig:subfig:b} 
    \includegraphics[width=40mm]{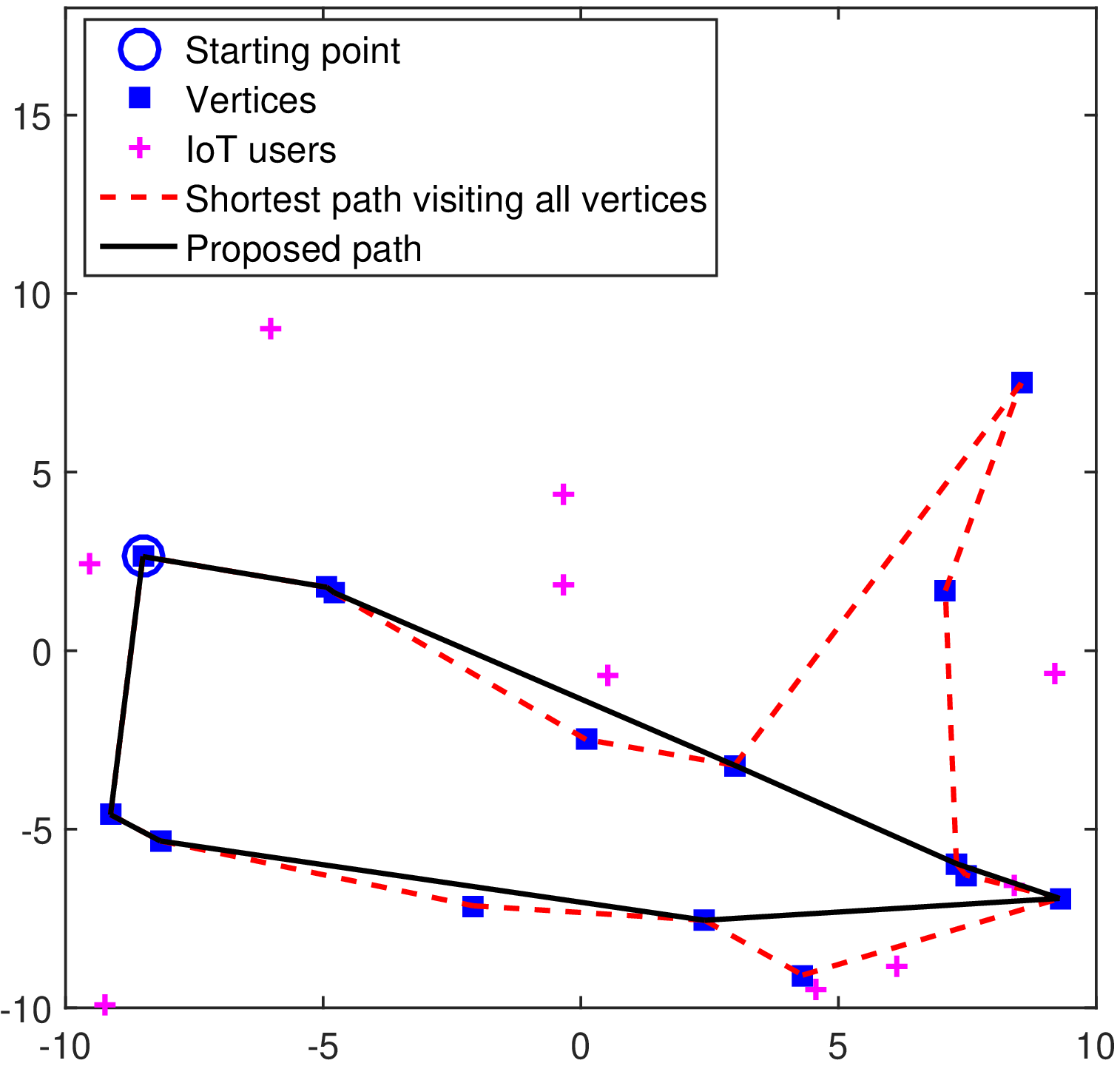}}
  \caption{(a) The proposed path with $K=10$ and $M=15$ when noise power $N_0=-90$ $\mathrm{dBm}$; (b) The proposed path with $K=10$ and $M=15$ when noise power $N_0=-60$ $\mathrm{dBm}$.
}
  \label{fig:subfig} 
\end{figure}

Next, we focus on the energy management performance of Algorithm 1.
In particular, the case of $K=10$ with $M=15$ is simulated, and the total energy consumption versus the noise power $N_0$ is shown in Fig. 2b.
It can be seen that if the noise power is large, by allowing the UGV to visit all the vertices, it is possible to achieve a significantly lower energy consumption compared to the case of no UGV movement.
However, this conclusion does not hold in the small noise power regime, which indicates that moving is not always beneficial.
Fortunately, the proposed Algorithm 1 can automatically determine whether to move and how far to move.
For example, if the noise power is extremely small (e.g., $-120~\mathrm{dBm}$), the UGV could easily collect the data from IoT users at the starting point.
In such a case, the proposed Algorithm 1 would fix the UGV at the starting point.
This can be seen from Fig. 2b at $N_0=-120~\mathrm{dBm}$, in which Algorithm 1 leads to the same performance as the case of no UGV movement.
However, if the noise power is increased to a medium value (e.g., $-90$ $\mathrm{dBm}$), the total energy is reduced by allowing the UGV to move (with the moving path shown in Fig. 3a).
On the other hand, if the noise power is large (e.g., $-60$ $\mathrm{dBm}$), the energy for data collection would be high for far-away users.
Therefore, the UGV should spend more motion energy to get closer to IoT users.
This is the case shown in Fig. 3b.
But no matter which case happens, the proposed algorithm adaptively finds the best trade-off between spending energy on moving versus on communication, and therefore achieves the minimum energy consumption for all the simulated values of $N_0$ as shown in Fig. 2b.

\section{Conclusions}

This paper studied a UGV-based backscatter data collection system, with an integrated graph mobility model and backscatter communication model.
The joint mobility management and power allocation problem was formulated with the aim of energy minimization subject to communication QoS constraints and mobility graph structure constraints.
An algorithm that automatically balances the trade-off between spending energy on moving and on communication was proposed.
Simulation results showed that the proposed algorithm could significantly save energy consumption compared to the scheme with no UGV movement and the scheme with a fixed moving path.

\appendices

\end{document}